%% file: QTcontrol_finitebias.tex
\documentclass[a4paper,12pt]{iopart}

\usepackage{graphicx}
\usepackage{amssymb}
 \usepackage{bm}
\usepackage{epsfig}
\usepackage{hyperref}
\hypersetup{breaklinks}
\usepackage{bbm}
\usepackage{xspace}

\newcommand{\rb}[1]{\left( #1 \right)}

\newcommand{\beq}{\begin{eqnarray}}
\newcommand{\eeq}{\end{eqnarray}}
\newcommand{\eq}[1]{Eq.~(\ref{#1})}
\newcommand{\fig}[1]{Fig.~\ref{#1}}
\newcommand{\secref}[1]{Sec.~\ref{#1}}

\newcommand{\citer}[1]{{Ref.\,\cite{#1}}}

 \usepackage[normalem]{ulem}
 \usepackage{color}

\begin{document}
\title[Quantum transport control: amplification, filtering and switching at finite bias]{Coherent control in quantum transport: amplification, filtering and switching at finite bias}
\author{Rajpal Thethi}
\address{
  Department of Physics and Mathematics,
  University of Hull,
  Kingston-upon-Hull,
  HU6 7RX,
  United Kingdom
}

\author{Clive Emary}
\address{
 Joint Quantum Centre (JQC) Durham-Newcastle,
 School of Mathematics and Statistics,
 Newcastle University,
 Newcastle-upon-Tyne,
 NE1 7RU,
 United Kingdom
}

\date{\today}
\begin{abstract}   
   We consider coherent feedback control of quantum transport and focus on the application of simple controllers and the effects of a finite bias voltage.
   We show that simple single-parameter controllers can give rise to a range of useful effects such as amplification of changes in plant transmission, increased resolution of energy filtration, and the detection of differences between otherwise indistinguishable plants.
   We explore how these effects are impacted by the phase-averaging effects associated with finite bias and identify important voltage scales for the maintenance of the functionalities achieved through feedback control.
\end{abstract}
\maketitle

\section{Introduction \label{SEC:Intro}}

Quantum transport studies the motion of electrons in structures small enough, and at temperatures low enough, that the quantum-mechanical nature of the electron plays a dominant role in determining their behaviour \cite{Nazarov2009}. Such studies are not only of fundamental interest,  but also point towards future quantum-technological developments in fields such as electronics and computation.  Unlocking such developments, however, requires that we learn to understand and control quantum effects in transport systems.

In 2010, Brandes \cite{Brandes2010} first proposed the use of quantum feedback control \cite{Wiseman2009, Gough2012,Zhang2015} to manipulate the flow of transport electrons and described a feedback scheme to suppress fluctuations in the inherently stochastic flow of electrons through a quantum dot.  This work spawned  a number of theoretical proposals to use feedback to produce a range of interesting transport effects such as stabilisation of nonequillibrium quantum states \cite{Kiesslich2011,Poeltl2011,Kiesslich2012} and the realisation of a mesoscopic Maxwell's demon \cite{Schaller2011, Esposito2012,Strasberg2013}. These ideas were reviewed in \citer{Emary2016a}, and recently, a scheme very close to Brandes' noise-suppression proposal was realised in experiment \cite{Wagner2016}. 

Most of these developments have utilised measurement-based quantum control \cite{Wiseman1994,Wiseman2009}, where the quantum system of interest (the ``plant'') is measured and the results processed as the informational grist of a classical feedback loop.  
In contrast, in \textit{coherent quantum control} \cite{Lloyd2000}, no explicit measurement is made but rather the plant is connected to a quantum-mechanical controller and their interactions form an autonomous feedback loop that is quantum-mechanical and phase coherent. 
The main advantages of coherent feedback control over its measurement-based counterpart are held to be reduced noise and increased speed \cite{Zhang2011}.

In Refs.\,\cite{Emary2014a,Gough2014}, Emary and Gough considered the application of coherent control to quantum transport, taking their cue from the {\em quantum feedback networks} of Gough and James \cite{Gough2008,Gough2009,Gough2009a,James2008,Nurdin2009,Zhang2011,Zhang2012}.
In their approach, transport was described within the Landauer-B\"uttiker theory \cite{Blanter2000}, applicable to non-interacting phase-coherent electron systems, with  plant and controller modelled by scattering matrices.  By coupling these building blocks together in loop geometries, a form of coherent feedback is implemented, the analysis of which amounts to finding the composite scattering matrix of the plant-controller complex.

\citer{Emary2014a} considered  a specific feedback geometry consisting of a four-lead plant attached to a two-lead controller. In the current work, we consider a complementary arrangement with a two-lead plant and a four-lead controller. 
Unlike the unrestricted controllers of \citer{Emary2014a}, the ones we consider here have particularly simple structures, with just a single control parameter. In terms of an optical analogy, these controllers are equivalent to banks of identical beam-splitters with the beamsplitting angle the control parameter.
We show here that, even with these highly-constrained controllers, interesting and useful actions of the feedback loop can be produced.
This is because the feedback loops exhibit strongly nonlinear relationships between the plant transmission and the transmission of the feedback circuit.  
We describe several applications of this effect: the first is to amplify differences in plant transmission originating from changes in an external parameter; the second is to increase the precision of a quantum dot energy filter; and the third is to distinguish between plants that are otherwise indistinguishable.

In contrast to previous works \cite{Emary2014a,Gough2014}, we consider here the performance of these feedback schemes at \textit{finite bias}. This is important because, at finite bias, electrons have a range of energies and thus pick us different phases as they travel around the feedback loop, and this affects the way they interfere.  Thus, we expect finite bias to lead to phase averaging \cite{Marquardt2004a,Chung2005,Haack2011}, which will, to some extent, undo the gains made by the coherent control.  We investigate the impact of this phase averaging both in general and on our applications and identify the voltage scale(s) over which feedback-enhanced functionalities can be maintained undiminished.

This paper is organised as follows. \secref{SEC:FB} introduces our feedback geometry and transport observables.  In \secref{SEC:phase} we address the phases acquired by the electrons as they travel through the feedback loop, and in \secref{SEC:controllers} we introduce our simple controllers and study the resultant feedback scattering matrices. 
We then turn to applications. \secref{SEC:amp} discusses the amplification of differences in the transmission of a single-channel plant caused by the variation of an external parameter.  Detailed account is given of the performance of this set up at both zero and finite bias. \secref{SEC:filter} then considers the plant to be a quantum dot, modelled as a single-resonant level, and considers the use of feedback to modify both the position and width of its resonance. Again, the effect of finite bias is discussed.  Our final application, discussed in \secref{SEC:multi}, considers a multi-channel system where feedback is employed to distinguish between two plant matrices that have identical transmission probabilities but different phases.  We conclude with the discussions in \secref{SEC:discussions}, including parameter estimations for realisation in quantum Hall edge channels.

\section{Feedback Geometry\label{SEC:FB}}

\begin{figure}[tb]
  \begin{center}
  \includegraphics[width=0.55\columnwidth,clip=true]{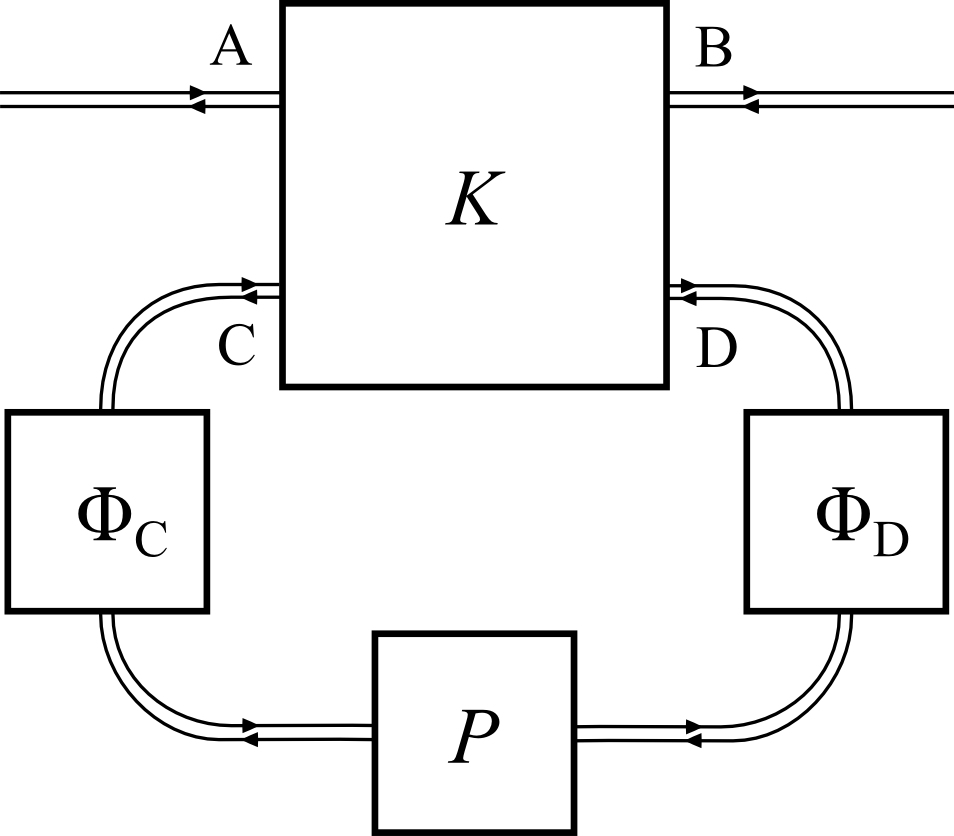}
  \end{center}
  \caption{
    Sketch of the feedback geometry considered here. Circuit blocks are labelled with the corresponding scattering matrix: $P$ is the plant, $K$ is the controller, and blocks $\Phi_\mathrm{C}$ and $\Phi_\mathrm{D}$, describe the phases accumulated by the electrons as they travel between plant and controller (in internal leads C and D). Conduction properties of the resultant feedback circuit are measured between leads A and B.
    \label{FIG:sketchFULL}
  }
\end{figure}

%
We consider the geometry sketched in \fig{FIG:sketchFULL}, in which a two-lead plant is connected in a loop geometry with a four-lead controller.  
This is similar to that considered in \citer{Emary2014a} but with the role of plant and controller reversed. Leads A and B are external leads, across which the transport properties of the plant-controller complex will be calculated.  Leads C and D are internal to the feedback loop. We assume that all leads are bidirectional and support $N$ transverse modes in the relevant energy range. 
Within the Landauer-B\"uttiker approach \cite{Blanter2000}, the plant is represented by the $2N\times2N$-dimensional scattering matrix $P(E,\lambda)$ which, in principle, depends on the energy $E$ at which scattering occurs as well as on some external parameter $\lambda$.  In terms of vectors $b^\mathrm{in}(E)$ and $b^\mathrm{out}(E)$ containing the appropriate annihilation operators of the leads,  the plant scattering matrix acts as
$
  {b}^\mathrm{out}(E) = P(E,\lambda) \;  b^\mathrm{in}(E)
$.
In terms of transmission and reflection blocks, $P = P(E,\lambda)$ has the form
\beq
  P  
  =
  \rb{
    \begin{array}{cc}
      r_P & t_P'
      \\
      t_P & r_P'
    \end{array}
  }
  \label{EQ:Psubblocks}
  .
\eeq
The control matrix is $4N\times4N$ dimensional and has block structure
\beq
  K = 
  \rb{
    \begin{array}{cc}
      K_\mathrm{I} & K_\mathrm{II} \\
      K_\mathrm{III} & K_\mathrm{IV}
    \end{array}
  }
  \label{EQ:Kfull}
  ,
\eeq
with 
\beq
  K_\mathrm{I} = 
  \rb{
    \begin{array}{cc}
      K_{\mathrm{AA}} & K_{\mathrm{AB}} \\
      K_{\mathrm{BA}} & K_{\mathrm{BB}}
    \end{array}
  }
  ;\quad
  K_\mathrm{II} = 
  \rb{
    \begin{array}{cc}
      K_{\mathrm{AC}} & K_{\mathrm{AD}} \\
      K_{\mathrm{BC}} & K_{\mathrm{BD}}
    \end{array}
  }
  ;
  \nonumber\\
  K_\mathrm{III} = 
  \rb{
    \begin{array}{cc}
      K_{\mathrm{CA}} & K_{\mathrm{CB}} \\
      K_{\mathrm{DA}} & K_{\mathrm{DB}}
    \end{array}
  }
  ;\quad
  K_\mathrm{IV} = 
  \rb{
    \begin{array}{cc}
      K_{\mathrm{CC}} & K_{\mathrm{CD}} \\
      K_{\mathrm{DC}} & K_{\mathrm{DD}}
    \end{array}
  }
  ,
\eeq
where submatrix $K_{YX}$ describes the scattering from lead $X$ to lead $Y$.

The final elements in \fig{FIG:sketchFULL} are the ``phase blocks'' labelled $\Phi_\mathrm{C}$ and $\Phi_\mathrm{D}$. These are scattering matrices that account for the phases accumulated by the electrons in travelling between plant and controller.
The scattering matrices read
\beq
  \Phi_\mathrm{C} = 
  \rb{
    \begin{array}{cc}
      0 & e^{i {\beta}_\mathrm{C}}
      \\
      e^{i {\alpha}_\mathrm{C}} & 0
    \end{array}
  }
  ;\qquad  
  \Phi_\mathrm{D} = 
  \rb{
    \begin{array}{cc}
      0 & e^{i {\alpha}_\mathrm{D}}
      \\
      e^{i {\beta}_\mathrm{D}} & 0
    \end{array}
  }
\eeq
where ${\alpha}_{\mathrm{C},\mathrm{D}}$ and  ${\beta}_{\mathrm{C},\mathrm{D}}$ are diagonal matrices of the various phases.  Whilst not strictly part of the plant, it behoves us for the formal development to combine these phases with the plant matrix. Utilising the bidirectional series-product $\Diamond$ \cite{Emary2014a}, we define
\beq
  P_\Phi(E,\lambda) &=& \Phi_\mathrm{C}(E) \Diamond  P(E,\lambda) \Diamond \Phi_\mathrm{D}(E) 
  \nonumber\\
  &=&
  \rb{
    \begin{array}{cc}
      e^{i {\beta}_\mathrm{C}} r_P e^{i {\alpha}_\mathrm{C}} 
        &  e^{i {\beta}_\mathrm{C}} t_P' e^{i {\alpha}_\mathrm{D}}
      \\
      e^{i {\beta}_\mathrm{D}} t_P e^{i {\alpha}_\mathrm{C}} 
        &  e^{i {\beta}_\mathrm{D}} r_P' e^{i {\alpha}_\mathrm{D}}
    \end{array}
  }
  .
\eeq
This can then be written as the simple matrix product
$
  P_\Phi(E,\lambda) 
  = e^{i \bm{\beta}} \cdot P \cdot e^{i \bm{\alpha}}
$,
where $\bm{\alpha}$ and $\bm{\beta}$ are matrices containing all phases of that particular type along the diagonals, first those from lead C and then those from lead D.

Following the argument of \citer{Emary2014a}, the scattering matrix of the complete feedback system is
\beq
  \mathcal{S}
  =
  K_\mathrm{I}
  +
  K_\mathrm{II}
  \frac{1}{\mathbbm{1}_{2N}- P_\Phi K_\mathrm{IV}}
  P_\Phi K_\mathrm{III}
  \label{EQ:SFB1}
  ,
\eeq
with $\mathbbm{1}_{2N}$, a unit matrix of dimension $2N\times 2N$.  

\subsection{Finite-bias current and noise}
The conduction properties of \eq{EQ:SFB1} for various plant and controller form the subject of this work.  The main observable and control target we will be interested in is conductance 
$
  G \equiv  I /V
$, where $V>0$ is the applied bias between leads A and B and $I$ is the resulting current. We will express the conductance in terms of 
$
  G_0=2e^2 /h
$, the conductance quantum (spin degeneracy is assumed).
In the Landauer-B\"uttiker approach, the current is given by
\beq
  I = \frac{2e}{h}
  \sum_n
  \int dE  
  \,
  T_{n}(E) 
  \left[
    f_\mathrm{A}(E) - f_\mathrm{B}(E)
  \right]
  \label{EQ:currentfull}
  .
\eeq
Here
$
  f_\mathrm{A,B}(E) = \left[ 1+e^{(E-\mu_\mathrm{A,B})/(k_\mathrm{B}\mathcal{T})}\right]^{-1}
$
are the Fermi functions of leads A and B with common electronic reservoir temperature $\mathcal{T}$ and chemical potentials $\mu_\mathrm{A,B}$ such that the voltage across the sample is $V=(\mu_\mathrm{A}-\mu_\mathrm{B})/e$.  Furthermore, $T_{n}(E)$ are the transmission probabilities associated with matrix $\mathcal{S}$. These are obtained as the eigenvalues of the matrix $t^\dag  t$ where $t=t(E)$ is the transmission subblock of the total scattering matrix $\mathcal{S}(E)$ [compare \eq{EQ:Psubblocks}].
In the zero-temperature limit, $ \mathcal{T}\to 0$, the Fermi functions reduce to step functions and the current becomes
\beq
  I = \frac{2e}{h}
  \sum_n
  \int_{\mu_\mathrm{B}}^{\mu_\mathrm{A}} dE  
  \,
  T_{n}(E) 
  .
\eeq
We will also consider the zero-frequency shot noise which, in the same limit, is given by \cite{Blanter2000}
\beq
  S &=& G_0 \sum_n \int_{\mu_\mathrm{B}}^{\mu_\mathrm{A}} dE 
  \, T_{n}(E)[1-T_{n }(E)]
  ,
\eeq
as well as the corresponding Fano factor
$
  F= S/(eI)
$.

\section{Phase and phase averaging \label{SEC:phase}}

The phases contained in matrices $\bm{\alpha}$ and $\bm{\beta}$ will, in general, consist of both geometrical and dynamical contributions. 
If we assume a linear dispersion about a Fermi energy of $E=0$, the wave number of the electron in the $n$th subband of lead $\nu = \mathrm{C},\mathrm{D}$ reads
$
  k_{\nu n} = k_{\nu n}(0) + E/ (\hbar v_{\nu n})
$,
with $k_{\nu n}(0)$ the relevant Fermi wavenumber  and $v_{\nu n}$ the corresponding velocity. Travelling a distance $L_{\nu n}$ between plant and controller, the phase accumulated by an electron will be \cite{Chung2005}
\beq
  \alpha_{\nu n}(E) &=& 
\alpha_{\nu n} (0) + \frac{L_{\nu n}}{\hbar v_{\nu n}(0)} E
  ,
\eeq
where $\alpha_{\nu n} (0)$ contains both the geometric phase and the Fermi-edge contribution $k_{\nu n}(0)L_{\nu n}$. A similar relation holds for $\beta_{\nu n}(E) $.

For simplicity, we will consider the homogeneous case for these phases, i.e. that the phase accumulated is the same for each channel and each lead, and assume\footnote{This situation could, for example, be approached in transport through quantum Hall edge channels by changing the local potential landscape of the different interconnecting ``leads'' to make the accumulated phases match, as well as by considering high-energy electron transport in the outer edge channels \cite{Kataoka2016} such that sub-band offsets make up but a small contribution to the total energy of the electrons and the velocities are thus uniform.}
\beq
  \alpha_{\nu n}(E) = \beta_{\nu n}(E) = \alpha = \alpha_0 + \frac{E}{eV_\Phi}
  \label{EQ:alpha0in}
  .
\eeq
This form introduces a single voltage scale associated with phase effects
\beq
  V_\Phi \equiv \frac{\hbar v}{e L}
  ,
\eeq
where $v$ and $L$ are the homogeneous electron velocity and path length between plant and controller.  In this case, the plant matrix develops an overall energy-dependent phase
\beq
  P_\Phi(E,\lambda) &=& e^{2 i \alpha_0 + 2 i E/(eV_\Phi) } P(E,\lambda)
  .
\eeq

We can assess the effect of this energy-dependent phase by assuming that plant and control matrices are independent of $E$, except through this phase.
In this case, we can decompose the scattering matrix as
$
  \mathcal{S}(E) = \sum_{n=0}^\infty  \mathcal{S}^{(n)} e^{2 i n E/(eV_\Phi)}
$, and similarily for its transmission subblock 
$
  t(E) = \sum_{n=0}^\infty  t^{(n)} e^{2 i n  E/(eV_\Phi)}
$.
Evaluating the energy-integral in \eq{EQ:currentfull} \cite{Bevilacqua2013}, we obtain the current 
\beq
  I &=& \frac{e}{h} \sum_{n,m=0}^\infty
  t^{(n)} {t^{(m)}}^\dag
  (2 \pi k_\mathrm{B} \mathcal{T})
  \mathrm{csch}
  \left[
    2\pi (n-m) \frac{k_\mathrm{B} \mathcal{T}}{eV_\Phi}
  \right]
  \sin
  \left[
    \frac{(n-m) V}{ V_\Phi}
  \right]
  \label{EQ:Idephased}
  ,
\eeq
where we have assumed a symmetrical bias $\mu_A = -\mu_B=eV/2$ and where $k_\mathrm{B} \mathcal{T}$ is the thermal energy of the leads.
This form shows the importance of the parameter $V_\Phi$ as setting the voltage scale over which phase-averaging effects become important. For temperature $k_\mathrm{B} \mathcal{T}\gg eV_\Phi$, the off-diagonal elements are exponentially suppressed by thermal fluctuations. In the low temperature limit $k_\mathrm{B} \mathcal{T}\ll eV_\Phi$, we obtain the conductance
\beq
  G
  =
  G_0
  \sum_{n,m=0}^\infty
  t^{(n)} {t^{(m)}}^\dag
  \mathrm{sinc}
  \left[
    \frac{(n-m) V}{ V_\Phi}
  \right]
  \label{EQ:Gdephased}
  .
\eeq
This form shows that for $V \ll V_\Phi$, the conductance reduces to its coherent limit and that the leading-order correction to this behaviour scales like $\rb{V/ V_\Phi}^2$.  At higher voltages, the conductance for each value of $n-m$ shows ``lobes'' as a function of voltage, similar to those predicted for the Mach-Zehnder interferometer \cite{Chung2005}. Here, however, since the electrons can make  multiple round trips in the loop, the overall conductance is the superposition of signals with different periods $(n-m)V/(2\pi V_\Phi)$.
A similar analysis can be performed for the noise, but the results are similar and we do not show it here.

\section{Simple controllers and switching \label{SEC:controllers}}

\begin{figure}[tb]
  \begin{center}
  \includegraphics[width=\columnwidth,clip=true]{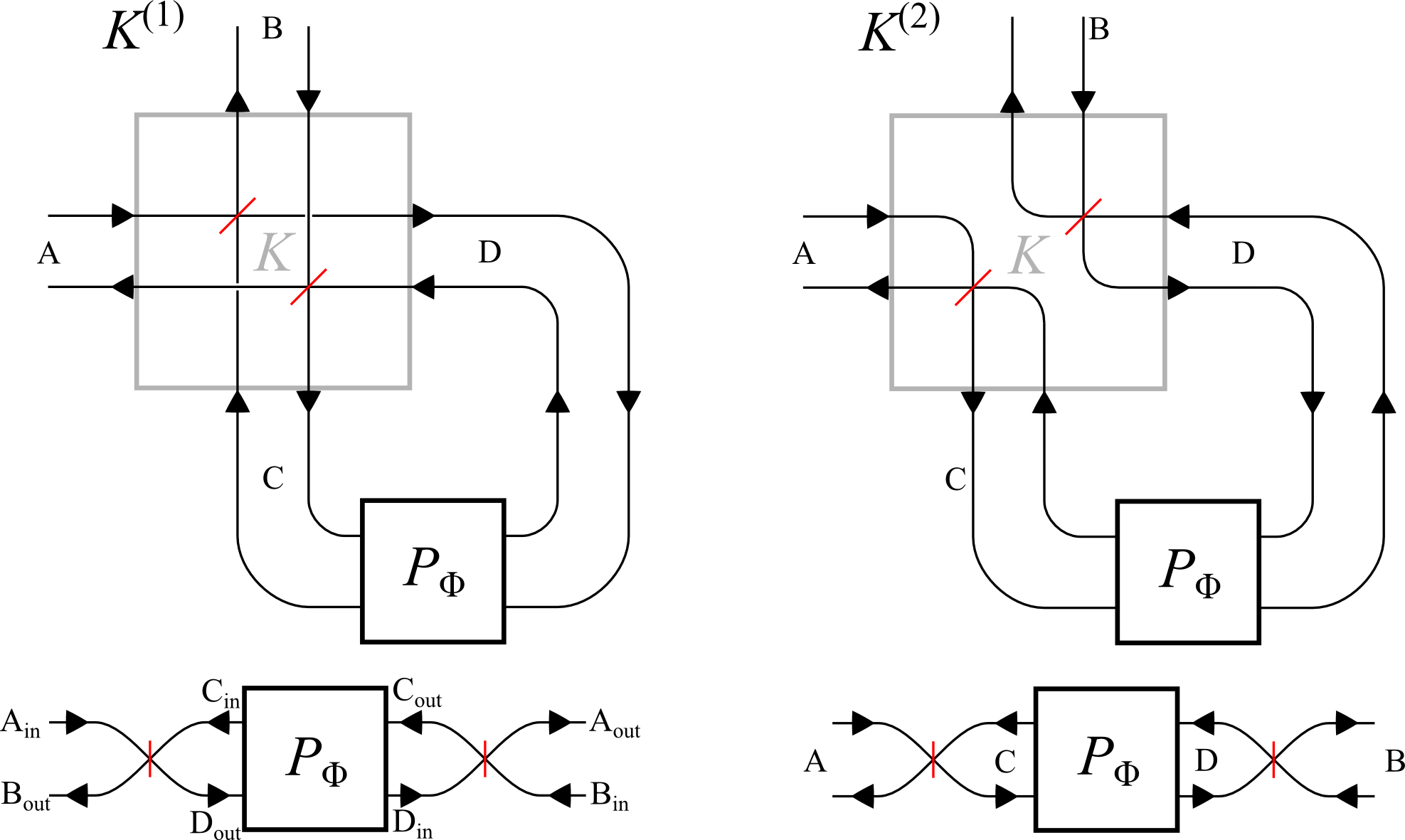}
  \end{center}
  \caption{
    Sketches of feedback loops with the two simple controllers, $K^{(1)}$ on the left and $K^{(2)}$ on the right, in the single-channel case.
    The action of each controller (large grey square) is decomposed in terms of two ``beamsplitters'', each represented by a short red line.
    The bottom two sketches show the circuits ``unfolded'' such that the action of the beamsplitters in each case is clear.
    In the $K^\mathrm{(1)}$ case, both beamsplitters scatter between all four leads. 
    In the $K^\mathrm{(2)}$ case, one beamsplitter acts between leads AC and the other between BD.  This second feedback loop can therefore be decomposed as the series combination of two beamsplitters with the plant in the middle.
    \label{FIG:sketchSIMPLE}
  }
\end{figure}

In \citer{Emary2014a} the notion of ``ideal control'' was discussed where, with free choice of control matrix, the scattering matrix for the joint plant-control system could be set arbitrarily regardless of the form of the plant matrix. For this to be possible, the control matrix needs to have dimension equal to or exceeding that of the plant matrix.  In this light it is clear then that without further restriction,  the 4-lead controller considered here will enable ideal control since its dimension is twice that of plant matrix $P$.

Here, however, we are interested in a different problem, viz., what can be achieved when we are restricted to particularly simple controllers.  We consider two such controllers, with scattering matrices $K^{(1)}$ and $K^{(2)}$, each of which just has a single control parameter, $\theta$, that controls the feedback ``strength''.  The controllers in question consist of a set of identical 4-port ``beamsplitters'' that preserve the channel index.  In a transport set-up, these would most readily be realised with quantum point contacts \cite{Wees1991}.  Sketches of these two controllers embedded in feedback loops are shown in \fig{FIG:sketchSIMPLE}.

The first controller $K^{(1)}$ scatters electrons between all four of its leads and has the scattering matrix
\beq
  K^{(1)} =
  \rb{
    \begin{array}{cc}
      \cos \theta X & \sin \theta X
      \\
       -\sin  \theta X' & \cos \theta X'
    \end{array}
  }
  ,
\eeq
where $X$ is an exchange matrix with elements $X_{ij} = \delta_{i,2N-j+1}$ and $X'$ is similar but with alternating signs $X'_{ij} = (-1)^{i}\delta_{i,2N-j+1}$. 
From \eq{EQ:SFB1}, the scattering matrix for the feedback loop with this controller reduces to 
\beq
  \mathcal{S}^{(1)} = X
  \frac{\cos\theta \, \mathbbm{1}_{2N} -  P_\Phi X'}
  {\mathbbm{1}_{2N} -  \cos\theta  P_\Phi X'}
  \label{EQ:SFBsimple1}
  .
\eeq
In the limit $\theta \to 0$, we obtain the feedback matrix  $\mathcal{S}^{(1)} \to X$; whereas for $\theta \to \pi/2$, we have $\mathcal{S}^{(1)} \to -X P_\Phi X'$.  Thus, varying the feedback parameter $\theta$ we can steer the system from a purely transmissive one, whose properties are independent of the plant, to a system whose scattering matrix is linear in the plant matrix. 
Exceptions to this generic behaviour occur when the denominator in \eq{EQ:SFBsimple1} is singular. For example, consider the plant matrix $P_\Phi = e^{2i\alpha_0} \mathbbm{1}_{2N}$, which represents a reflection of all channels with a $2\alpha_0$ phase accumulation.  In this case, the determinant of denominator in \eq{EQ:SFBsimple1} with 
$\theta \to 0$ reads 
$\mathrm{Det}\rb{\mathbbm{1}_{2N} -  e^{2i\alpha_0}_\Phi X'} = \rb{1+e^{i 4\alpha_0}}^{N}$.  This is zero for $\alpha_0 = (2m+1)\pi/4;~~m=0,\pm1,\pm2,\ldots$ and \eq{EQ:SFBsimple1} can not be used directly. However, re-evaluating $\mathcal{S}^{(1)}$ without the series resummation for the special case of $P_\Phi = i \mathbbm{1}_{2N}$ gives the feedback matrix $\mathcal{S}^{(1)} = -i XX' = \mathrm{Diag}[-i,i,-i,\ldots]$ for all values of the feedback parameter $\theta$.  In this case, then, the system is completely reflective.

We will make use of this behaviour to give a circuit whose conduction properties depend abruptly on the plant matrix. With $\theta$ small, most plant matrices will give a feedback circuit that is highly transmissive since, for $\theta \to 0$, we have $\mathcal{S}^{(1)} \to X$. However, when the plant matrix resembles $P_\Phi = i \mathbbm{1}_{2N}$, the behaviour of the feedback circuit will switch to the singular limit $\mathcal{S}^{(1)} = -i XX'$, which is purely reflective. Thus, the behaviour of the feedback circuit will switch rapidly as the plant approaches a purely reflective state.

The control matrix of our second controller reads
\beq
  K^{(2)} =
  \rb{
    \begin{array}{cc}
      \cos \theta \mathbbm{1}_{2N} & \sin \theta \mathbbm{1}_{2N}
      \\
       -\sin  \theta \mathbbm{1}_{2N} & \cos \theta \mathbbm{1}_{2N}
    \end{array}
  }
  .
\eeq
This induces scattering only between leads $A$ and $C$, and between $B$ and $D$.  It thus effectively acts as two sets of beamsplitters, one on either side of the plant, see \fig{FIG:sketchSIMPLE} and use of this controller therefore resembles enclosing the plant in what, in optical terms, would be something like a generalised Fabry-P\'{e}rot resonator \cite{Ismail2016}. 

The feedback matrix for the feedback loop with $K^{(2)}$ reads
\beq
  \mathcal{S}^{(2)} = 
  \frac{\cos\theta \mathbbm{1}_{2N} - P_\Phi}
  {\mathbbm{1}_{2N} - \cos \theta P_\Phi}
  .
\eeq
For $\theta \to 0$, the feedback matrix becomes $\mathcal{S}^{(2)} \to \mathbbm{1}_{2N}$, whereas for $\theta \to \pi/2$, we obtain $\mathcal{S}^{(2)} \to -P_\Phi$. Thus varying $\theta$ changes the circuit from complete reflection to being the same as the plant except for an overall phase.  
In a similar fashion to the above, there is an exception to this behaviour when the denominator is singular. When, for example,  $P =  i X'$ we obtain $\mathcal{S}^{(2)} = -i X'$ instead. Thus, this set-up shows an abrupt switching behaviour with  $\theta \to 0$ from reflection for most scattering matrices to complete transmission for e.g. $P_\Phi \to  i X'$.

\section{Signal amplification \label{SEC:amp}}

\begin{figure}[tb]
  \begin{center}
    \includegraphics[width=1\columnwidth,clip=true]{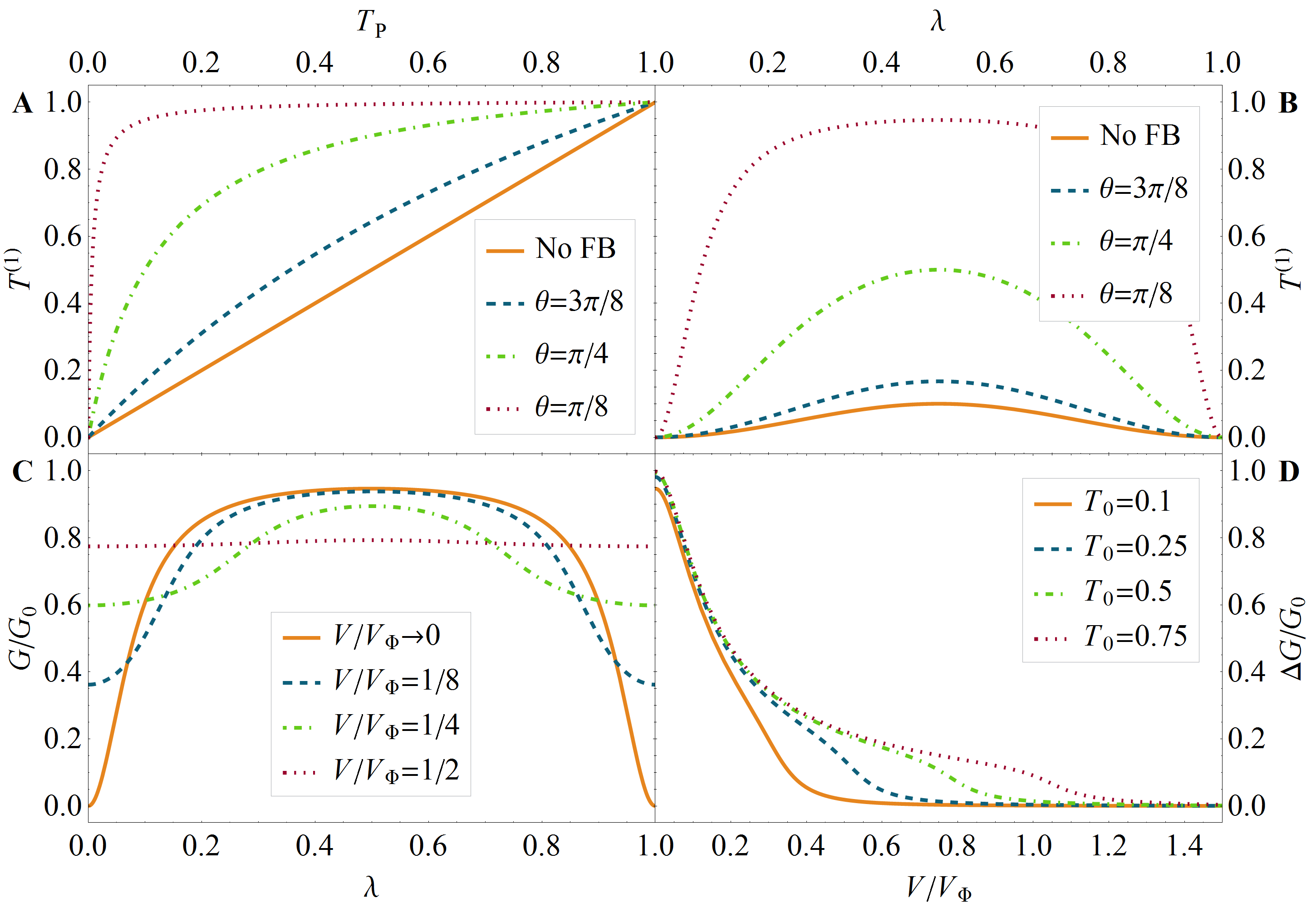}
  \end{center}
    \caption{
      Signal amplification with the feedback controller $K^{(1)}$.
      {\bf A.}
      Transmission of the feedback circuit, $T^{(1)}$, as a function of the plant transmission, $T_\mathrm{P}$, with plant matrix \eq{EQ:P(T)} for $E=0$.  Changing the feedback control parameter $\theta$ changes the degree of nonlinearity and hence the scope for amplifying changes in plant transmission.
      {\bf B.} Transmission $T^{(1)}$ when the plant transmission is a sinusoidal function of parameter $\lambda$, see \eq{EQ:Tamplify}.  With $T_0=0.1$ the amplitude of these oscillations is small. However, with increasing $\cos\theta$, the nonlinear response of the feedback circuit leads to an amplification of their magnitude.
      {\bf C.} The conductance on this circuit as a function of $\lambda$ for various voltages.  The feedback angle here is $\theta=\pi/8$ such that amplification is strong. As voltage increases, the high-amplitude oscillations become smoothed out due to phase averaging.
      {\bf D.} The peak-to-peak conductance $\Delta G$ of the oscillations as a function of voltage ($\theta=\pi/8$).  This shows that the signal amplification drops by half over a scale of $V/V_\Phi \approx 0.2$. For voltages lower than this, amplification is robust with the results largely independent of $T_0$. 
      In all cases, the fixed phase angle was $\alpha_0 = \pi/4$.
      \label{FIG:K1cond}
    }
\end{figure}

Our first application of this formalism is to look at how the feedback loop can be used to amplify changes in the transmission of a single-channel plant. Let us  consider a plant matrix 
\beq
  P_\Phi
  &=&
  e^{2 i \alpha_0 + 2i E/(eV_\Phi) }
  \rb{
    \begin{array}{cc}
  	\sqrt{1-T_\mathrm{P}} 	&	\sqrt{T_\mathrm{P}}	\\
     -\sqrt{T_\mathrm{P}} 
      & \sqrt{1-T_\mathrm{P}}
    \end{array}
  }
  \label{EQ:P(T)}
  ,
\eeq
where $T_\mathrm{P} \in [0,1]$ is the plant transmission probability. We set  $\alpha_0 = \pi / 4$ in order to trigger the singular limit discussed above, and will initially consider the $E=0$ case. 
The transmission of the feedback circuit for this plant with control matrix $K^{(1)}$ as a function of the plant transmission $T_\mathrm{P}$ is
\beq
 T^{(1)}[\,T_\mathrm{P}] = \frac{T_\mathrm{P} \left(3+\cos2\theta\right)^2}{4\left( 4T_\mathrm{P}\cos^2\theta + \sin^4 \theta\right)}
 \label{EQ:T1explicit}
 ,
\eeq
and this result in plotted in \fig{FIG:K1cond} A. 
%
%
We see that the feedback-circuit transmission is a monotonically-increasing function of the plant transmission with a nonlinearity that increases with increasing $\cos\theta$. Indeed for small $\theta$, the transmission is approximately unity across most of the range of $T_\mathrm{P}$, and only when $T_\mathrm{P}$ approaches zero, does the behaviour switch and the transmission rapidly drop to zero.
For small $T_\mathrm{P}$, we find
$
  T^{(1)} \approx \rb{1-2\csc^2\theta}^2 T_\mathrm{P}
$, which is valid for $T^{(1)}\ll 1$. The quantity $\eta =\rb{1-2\csc^2\theta}^2$ thus represents the small-signal gain for the transmission in this circuit.  For $\theta = \pi/4$ and $\theta = \pi/8$, this evaluates as approximately 9 and 160 respectively.

To investigate the effects of applied bias on this amplification, we will consider plant matrix \eq{EQ:P(T)} with a plant transmission $T_\mathrm{P}$ independent of energy but depending sinusoidally on an external parameter $\lambda$:
\beq
  T_\mathrm{P} = T_\mathrm{P}(\lambda) = T_{0} \sin^2 \rb{\pi \lambda}
  \label{EQ:Tamplify}
  ,
\eeq
with $T_0$ the maximum transmission of the plant and hence the signal amplitude.  A plant of this form might, for example, arise from an interferometer geometry with $\lambda$ proportional to some external magnetic field or difference in path-length.  We shall assume that $T_0 \ll 1$ and that the aim of the feedback loop is to increase the amplitude of the oscillations. We set  $\alpha_0 = \pi / 4$ again and maintain the energy-dependence of the phase.

\fig{FIG:K1cond}B shows the transmission of the feedback circuit as a function of control parameter $\lambda$ with $T_0=0.1$, and the amplification of the 
original signal is apparent. For small $\cos \theta$, the response is approximately linear with the gain $\eta$ above. For larger $\cos \theta$, the nonlinearity of the response is manifest and we see distortion of the original sinusoid.
\fig{FIG:K1cond}C illustrates what happens at finite bias.  If we had only the plant matrix between our contacts, the conductance oscillations would be unaffected by the increase in bias. However, in the feedback circuit, the amplified oscillations are subject to phase-averaging and they thus wash out with increasing bias.
This point is illustrated further in \fig{FIG:K1cond}D which shows the peak-to-peak conductance $\Delta G = G_\mathrm{max}-G_\mathrm{min}$ of the feedback circuit as a function of the applied bias for $\theta = \pi/8$.
At small bias, $V \to 0$, the conductance difference $\Delta G$ is almost one. However, by the time the voltage reaches $V/V_\Phi\approx 0.2$, it has been reduced to $\Delta G \approx 0.5$  and it then trails off further.  This highlights the importance of the role of parameter $V_\Phi$ is determining the voltage scale over which phase-averaging effects act.   That the conductance disappears over a scale of $V/V_\Phi\approx 0.2$ and not $V/V_\Phi\approx \pi$ is indicative that more than the lowest voltage-dependent component in \eq{EQ:Gdephased} is contributing to the phase averaging here, and hence multiple round-trips are being made by the electrons in the feedback loop.
These results also show a degree of robustness with respect to variations in parameter $T_0$, which is to say that, providing the voltage is below the threshold at which significant phase-averaging occurs, the peak-to-peak conductance remains the same for a wide range of $T_0$ values. This arises from  the saturation of upper bound for the transmission.  Above $V/V_\Phi\approx 0.2$, however, dephasing is strong and this robustness disappears, resulting in a peak-to-peak conductance that very much depends on the plant transmission.

\begin{figure}[tb]
  \begin{center}
    \includegraphics[width=1\columnwidth,clip=true]{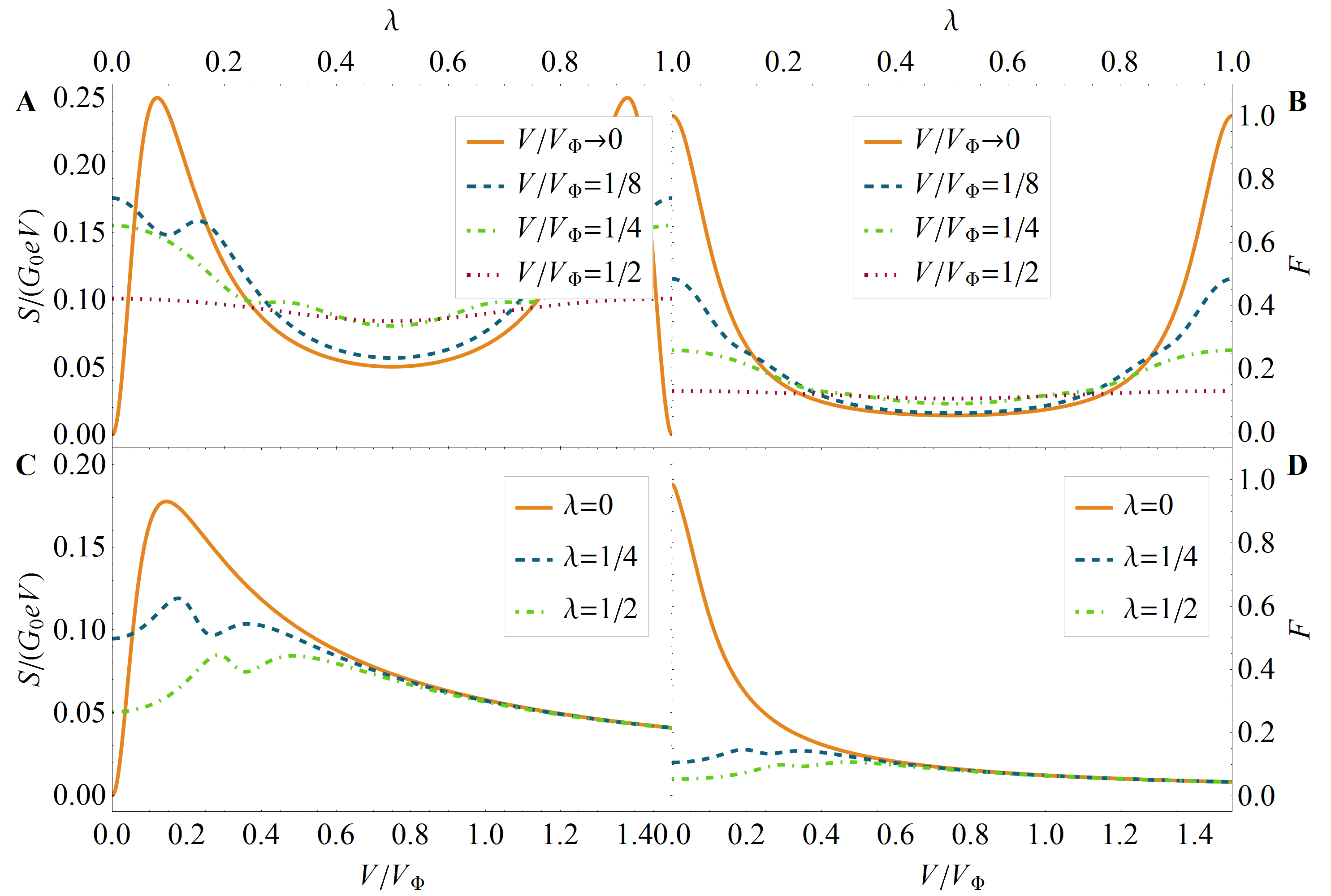}
  \end{center}
      \caption{
        Noise properties of the single-channel feedback circuit discussed in \fig{FIG:K1cond}. 
        Panels {\bf A.} and {\bf B.} show the shotnose and Fano factor as a function of external parameter $\lambda$ at various applied bias.
        Panels {\bf C.} and {\bf D.} show the shotnose and Fano factor as a function of bias for select values of $\lambda$.
	At low voltage, and away from from the endpoints $\lambda = 0,1$, the feedback gives rise to a strong suppression of the Fano factor. As voltage increases, the $\lambda$-dependence is washed out and suppression spreads to envelop the complete range of $\lambda$.
	Parameters as in \fig{FIG:K1cond}C unless otherwise stated.
	\label{FIG:K1noise}
      }
\end{figure}

In \fig{FIG:K1noise} we consider the noise properties of this system.  In the limit $V \to 0$ the noise and Fano factor for this single-channel model are simply proportional to $T^{(1)}(1-T^{(1)})$  and $1-T^{(1)}$ respectively.  For small $T_0$ without feedback, the Fano factor is therefore close to one through the whole range of $\lambda$. The application of feedback, however, reduces the noise relative to the current, leading to a dramatic reduction of the Fano factor, most markedly at $\lambda=1/2$. This behaviour can be seen in Panels A and B of \fig{FIG:K1noise}.  These figures also show the noise and Fano factor at finite bias and both show a smearing out of the $\lambda$-dependent behaviour as voltage increases.  What is remarkable is that, for the parameters considered here, the Fano factor in the large-voltage limit is close to 0.1 across the entire $\lambda$ range.  Thus, even in the strongly dephased classical limit, the feedback loop leads to a suppression of the Fano factor. This is particularly marked around $\lambda = 0$ and $\lambda = 1$ where the Fano factor is one without feedback.

\section{Energy filter \label{SEC:filter}}

We now consider the second controller $K^{(2)}$ and first consider it acting on a plant as in \eq{EQ:P(T)} with $\alpha_0=\pi/4$ and $E=0$.  The output transmission in this case $T^{(2)}$ can be related to the response of \eq{EQ:T1explicit} as 
\beq
  T^{(2)}[T_\mathrm{P}] = 1 - T^{(1)}[1-T_\mathrm{P}]
  .
\eeq
The result is thus a mirroring of the transmission in \fig{FIG:K1cond}A along the ``No FB'' line such that transmission $T^{(2)}$ is close to zero for most of its range with a rapid jump to $T^{(2)}\approx 1$ close to $T_\mathrm{P} = 1$.
The dominant suppression of transmission shown by this response can be utilised to sharpen features in the plant transmission.  To show this, we consider a single-channel plant whose transmission is dependent on energy (rather than on some external parameter as before), and take as example a single-level quantum dot with scattering matrix \cite{Buttiker1990}
\beq
  P_\Phi(E)=
  e^{ 2i \alpha(E)}
  \rb{\begin{array}{cc}
    1-\frac{i \Gamma}{ i \Gamma+E -E_r} 
  & -\frac{i  \Gamma}{ i \Gamma+E -E_r} \\
  -\frac{i  \Gamma}{ i \Gamma+E -E_r} 
  & 1-\frac{i \Gamma}{ i \Gamma+E -E_r}   \\
  \end{array} }
  .
\eeq
Here, $E_r$ is the energy of the resonant level and $\Gamma$ describes its width (we assume equal left and right barriers).  The transmission of this matrix is the Lorentzian
$
  T_\mathrm{P}(E) = \Gamma^2/\left[\Gamma^2+\left(E-E_r\right)^2\right]
$.
%
\begin{figure}[tb]
  \begin{center}
     \includegraphics[width=1\columnwidth,clip=true]{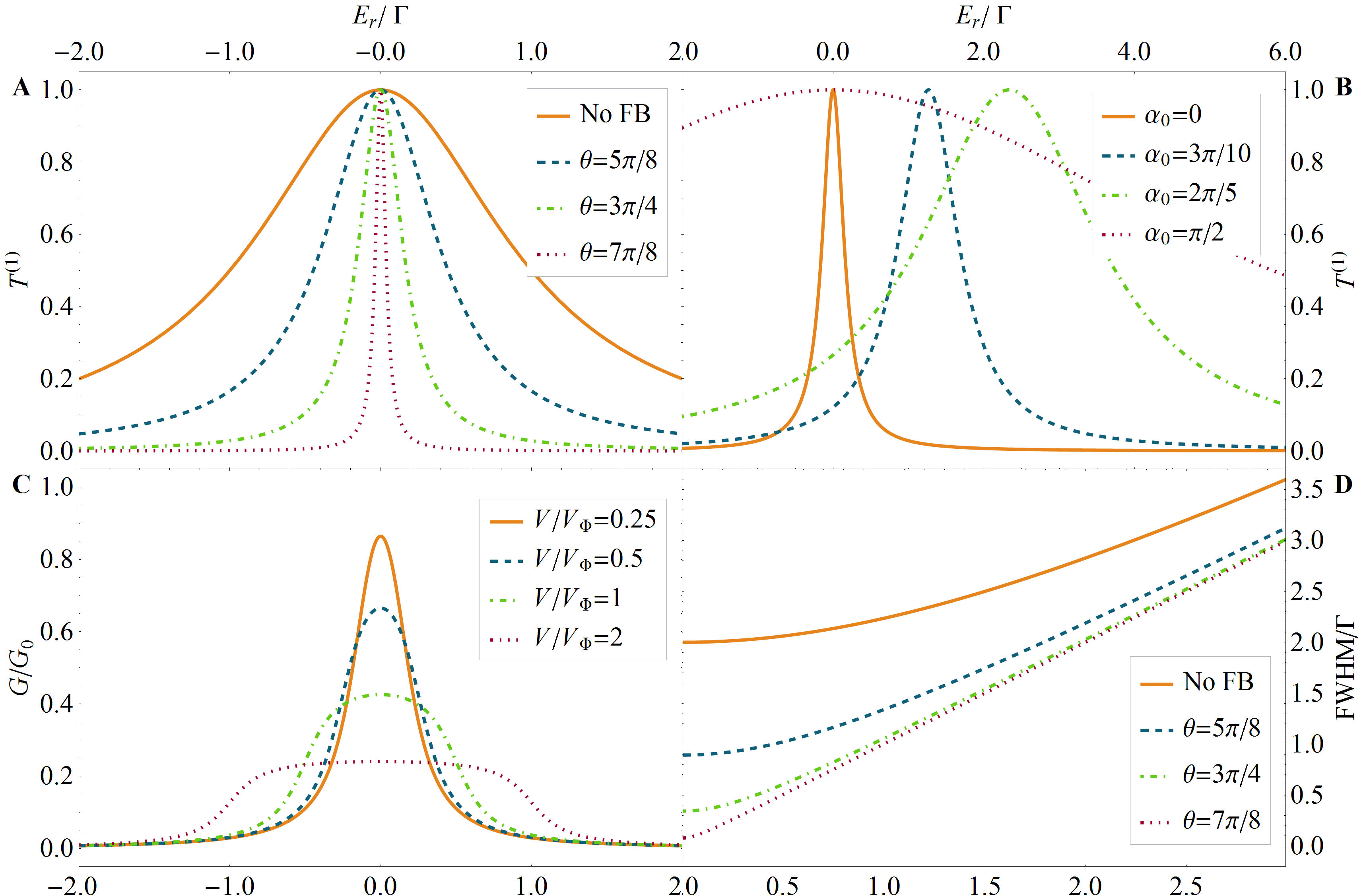}
  \end{center}
  \caption{
    Conductance properties of the feedback circuit with a single-level quantum dot plant and $K^{(2)}$ controller.
    {\bf A:} Without feedback, the transmission of the single-level quantum dot as a function of the resonance position $E_r$ is a Lorentzian with width parameter $\Gamma$.  With phase angle $\alpha_0=0$, changes in control angle $\theta$ reduce the width of the resonance.
    {\bf B:} Variations in the phase angle $\alpha_0$ can shift the transmission peak at the same time as changing its width. Here, the control angle was $\theta=3\pi/4$.
    {\bf C:} The conductance $G$ at finite bias.  As voltage is increases, the resonance broadens and flattens.  Here $\theta=3\pi/4$ and $\alpha_0=0$.
    {\bf D:} The FWHM of the conductance peak as a function of voltage for various control angle $\theta$.
    \label{FIG:QD}
  }
\end{figure}
%
Figure \ref{FIG:QD}A shows the transmission of the plant both without and with feedback (static phase $\alpha_0=0$).  Increasing feedback parameter $\theta > \pi/2$ causes the resonance to sharpen. This feedback scheme could therefore be used to increase the utility of the quantum dot as an energy-filter, e.g. \cite{Bhadrachalam2014,Tewari2016}. 
For values $\theta < \pi/2$, the feedback loop can be used to increase the width of the resonance (not shown).  
Figure \ref{FIG:QD}B shows that by changing the overall phase $\alpha_0$, not only the width, but also the position of the transmission resonance can be altered.
 
Choosing parameters such that the feedback gives a pure narrowing of the resonance, we now look at the effects of bias on this effect.  Figure \ref{FIG:QD}C shows the  conductance of the dot as a function of the resonance position with the centre of the bias window set at an energy $E=0$.
As bias increases, the resonant profile of the conductance peak flattens and broadens.  
Figure \ref{FIG:QD}D shows the FWHM of the conductance peak as a function of voltage. With no feedback and for $eV \ll\Gamma$, the FWHM is approximately constant as the bias window is still able to resolve the details of the resonance. For $eV\gg\Gamma$, however, the entire resonance fits inside the bias window such that that FWHM increases linearly with $V$. Somewhat surprisingly, we observe that this picture remains in tact under the application of feedback, with only the width of the original resonance being reduced. This indicates that the phase-averaging caused by the small finite bias does not significantly affect the resonance-narrowing effect of the feedback.

\section{Multi-channel switching \label{SEC:multi}}

We now extend our analysis to the multi-channel case and consider an $N$-channel plant that can switch discretely between two states, each with its own scattering matrix. We assume that the difference in the conduction properties of the two plant states is negligible and the role of the feedback circuit is to help distinguish between them.

We decompose our plant scattering matrices as \cite{Beenakker1997}
\beq
  P = 
  \rb{
    \begin{array}{cc}
      U & 0 \\ 0 & V
    \end{array}
  }
  \rb{
    \begin{array}{cc}
      -\sqrt{1-\mathbf{T}_P} & \sqrt{\mathbf{T}_P} \\ \sqrt{\mathbf{T}_P} & \sqrt{1-\mathbf{T}_P}
    \end{array}
  }
  \rb{
    \begin{array}{cc}
      U' & 0 \\ 0 & V'
    \end{array}
  }
  \label{EQ:SRMT}
  ,
\eeq
where $U$, $V$, $U'$, $V'$ are four $N \times N$ unitary matrices and
$\mathbf{T}_P = \mathrm{diag} (T_{P1} , T_{P2} , \ldots T_{PN})$ is an $N\times N$ diagonal matrix with the transmission eigenvalues on the diagonal.
We consider the extreme case for our problem, where all the transmission probabilities $T_{Pn}$ for the two plant matrices are homogeneous and identical to one another.  The two plant matrices thus differ only in the phase matrices $U$, $V$, etc. Since the transmission probabilities are the same for both matrices, the two plants alone have indistinguishable conductance properties. By enclosing the plants in a feedback loop, however, the phase matrices play a role, and thus lead to the possibility that the plants may be separated.

\begin{figure}[tb]
  \begin{center}
    \includegraphics[width=0.9\columnwidth,clip=true]{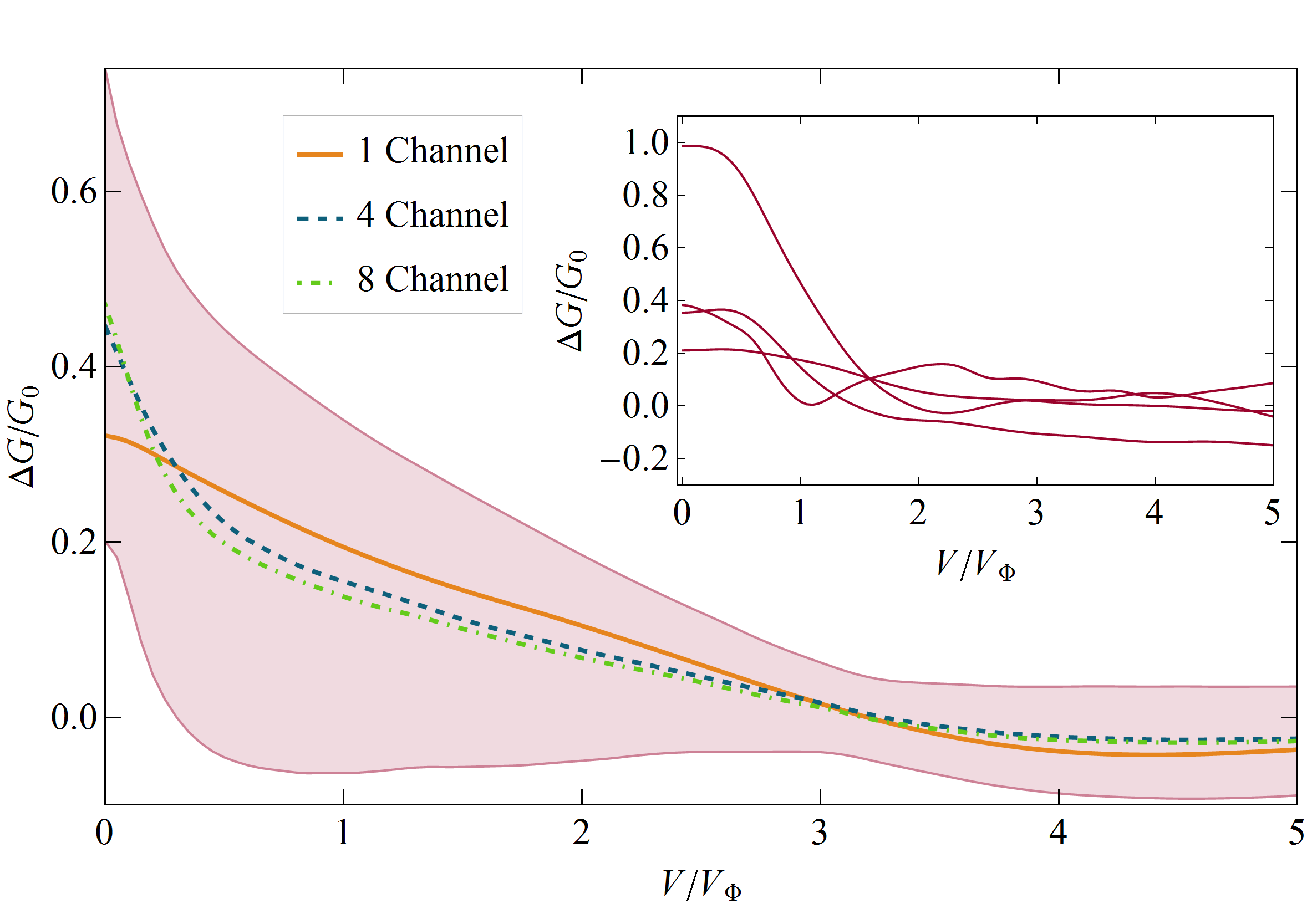}
  \end{center}
  \caption{
    \textbf{Main panel:}  The average conductance difference $\overline{\Delta G}$ between two plant matrices with identical transmission probabilities when placed in a feedback loop with controller $K^{(1)}$.  Results are shown as a function of voltage for $N=1,4,8$-channel matrices with the mean constructed over 300 random matrix pairs.  The pink region represent one standard deviation from the mean ($N=8$ case).
    \textbf{Inset:} Four individual instances of the conductance difference for $N=8$. Similar results were obtained for other values of $N$. 
    In all cases the transmission amplitudes of the plant matrices were all set to be $T_0=0.5$.
    \label{FIG:RMT}
  }
\end{figure}

To assess this effect, we consider a large set of plant matrices, generated randomly.  We generate a pair of matrices by choosing each of $U$, $V$, etc. from the circular unitary ensemble and keep all transmission probabilities equal and fixed at a value $T_0$.  We then calculate the conductance of each of these matrices when combined with controller $K^{(1)}$ with $\alpha_0 = 0$ and then vary the angle $\theta$ such that the difference in conductance $\Delta G$ at zero bias for any particular pair is maximised.  We repeat the procedure for 300 pairs of matrices of increasing size $N$, and look at the average value of $\Delta G$ for the set.  The results as a function of bias are outlined in \fig{FIG:RMT}.
%

Without feedback, $\Delta G$ would be zero by construction. At zero bias, however, we see that the average conductance difference $\overline{\Delta G}$ with the feedback loop is $\approx 0.5 G_0$.  This is then seen to saturate as the number of channels $N$ increases and thus, relative to overall conductance, the size of the control effect decreases like $1/N$. While a decrease with $N$ is unsurprising, since we have but a single control parameter irrespective of the size of plant we seek to control, it is noteworthy that the relative size of the control effect scales like $1/N$ and not $1/N^2$, as might be expected based on a parameter-counting argument.
Turning to finite bias, we see that the magnitude of $\overline{\Delta G}$ drops off with increasing $V$ due to phase averaging. Indeed $\overline{\Delta G}$ reaches zero when $V/V_\Phi \approx  \pi$, showing that it is the first off-diagonal contribution in \eq{EQ:Gdephased} that is dominant in the averaged feedback behaviour.
\fig{FIG:RMT} also makes clear that there is significant variation between the various instances of plant-matrix pairs. The standard deviation of our results for $N=8$ is shown in \fig{FIG:RMT}.  This shows a narrowing around the $V/V_\Phi \approx  \pi$, again indicating the importance of the first off-diagonal term.  The inset of this figure shows several results for individual plant-matrix pairs.  These display highly non-monotonic behaviour as functions of $V$, which also includes sign reversal. Nevertheless, the overall trend is towards a decreasing $\Delta G$ for large $V/V_\Phi > \pi$.

\section{Discussion \label{SEC:discussions}}

In this work we have considered coherent control of quantum transport within a scattering approach, and have focussed on the control effects achievable with two simple, generic controllers.  These controllers each give rise to a circuit whose transmission is a nonlinear function of plant transmission and this has been seen to be useful in a number of contexts.  We have described how this nonlinearity can be used to amplify changes in the plant transmission or detect changes where none would be visible in the transmission of the plant alone.  We have also seen how the feedback circuit can affect transmission as a function of energy, and in particular in enhancing the energy-filtration provided by a quantum dot. This range of application demonstrates that, even though we are far from the ``ideal control'' scenario open to unrestricted controllers \cite{Emary2014a}, even highly-constrained controllers can be of significant utility.

We have also considered how these schemes function at finite applied bias and have  identified the voltage scale $V_\Phi$ over which phase-averaging effects becomes significant.  In the amplifier application, the gain was cut by a half when the voltage reached $V/V_\Phi \approx 0.2$, and this we take to be a typical value for the point at which coherent control functionality is significantly degraded. The energy-filter example did not show any strong disruption at finite bias, but this was because the width of the resonance is masked at high bias anyway. Finally, in our multi-channel example, a complicated response as a function of voltage was observed, which reflects the non-trivial energy structure of circuits involving multiple round-trips. Interestingly, when such behaviour is averaged over a large number of plants, it is the phase-averaging properties of a single round-trip that dominates.

The most likely current physical context in which these ideas could be tested is in the edge-channel transport in the quantum Hall effect, where electronic analogues of optical beamsplitters are readily realised with quantum point contacts \cite{Wees1991}. A large range of interferometric experiments have been realised with such set-ups \cite{Oliver1999,Ji2003,Roulleau2008,McClure2009,Tewari2016}.
For typical edge-channel velocities of $v=10^5$ms$^{-1}$  \cite{McClure2009,Kataoka2016} and  interconnects $L=10$nm in length, we obtain a value for the phase-averaging scale $V_\Phi \sim 10$mV.
For $L = 10\mu$m, which is a typical Mach-Zehnder arm length \cite{Tewari2016} or coherence length at filling factor $\nu=2$ \cite{Roulleau2008}, we obtain $V_\Phi \sim 10\mu$V.
To avoid the negative phase-averaging impacts of finite bias on feedback-enhanced functionality, the operating bias should be kept significantly below these levels and this clearly suggests interconnects closer to the former value in length than the latter.

We have only considered the homogeneous case for the phases here.
For any particular geometry, inhomogeneous phases can be included without issue.  And whilst this will certainly complicate the specifics of the response, the same general principles are expected to  hold.  For a simple estimation of whether phase-averaging will impact the circuit, the largest value of the length-to-velocity ratio $L/v$ should be considered.

Finally, we note that the only incoherent process we have considered here is phase averaging due to finite bias.  In any physical realisation, other incoherent effects such as Coulomb interactions \cite{Levkivskyi2008,Taubert2011,Lunde2016,Nigg2016} and phonon emission \cite{Fletcher2013,Emary2016} will also contribute to degrade coherent feedback schemes.  Impact of these effects remains the subject of future work.

\ack{
  RT was funded by the University of Hull.
}

\section*{References}
\input{QTcontrol_finitebias.bbl}

\end{document}

%% file: QTcontrol_finitebias.bbl
\providecommand{\newblock}{}